\documentclass[a4paper,usenatbib]{mn2e}
\usepackage{natbib,epsfig}

\catcode`\@=11
\def\gta{\ifmmode{\,\mathrel{\mathpalette\@versim>\,}}
    \else{$\,\mathrel{\mathpalette\@versim>}\,$}\fi}
\def\lta{\ifmmode{\,\mathrel{\mathpalette\@versim<\,}}
    \else{$\,\mathrel{\mathpalette\@versim<}\,$}\fi}
\def\@versim#1#2{\lower 2.9truept \vbox{\baselineskip 0pt \lineskip
    0.5truept \ialign{$\m@th#1\hfil##\hfil$\crcr#2\crcr\sim\crcr}}}
\catcode`\@=12  
\newcommand{\beq}{\begin{equation}}
\newcommand{\eeq}{\end{equation}}
\def\figref#1{Fig.~\ref{#1}}
\def\b#1{{\bf#1}}
\def\vlos{v_\parallel}
\def\vperp{v_\perp}
\def\Flos{F_\parallel}

\def\kms{\,{\rm km}\,{\rm s}^{-1}}

\def\kpc{\,{\rm kpc}}

\def\d{{\rm d}}\def\D{{\rm D}}

\newcommand{\vect}[1]{{\bf #1}}
\newcommand{\vecthat}[1]{\hat\vect{#1}}
\newcommand{\eqref}[1]{(\ref{#1})}
\def\vot{\vect{v}_{0\perp}}

\title[Fitting orbits to tidal streams]
{Fitting orbits to tidal streams with proper motions}

\author[A. Eyre and J. Binney]{Andy Eyre and James  Binney\\
Rudolf Peierls Centre for Theoretical Physics, Keble Road, Oxford OX1 3NP, UK\\}

\begin{document}

\date{Draft, February 7, 2008}

\pagerange{\pageref{firstpage}--\pageref{lastpage}} \pubyear{2007}

\maketitle

\label{firstpage}

\begin{abstract}
The Galaxy's stellar halo seems to be a tangle  of disrupted
systems that have been  tidally stretched out into streams. Each stream approximately
delineates an orbit in the Galactic force-field. In the first paper in this
series we showed that  all six phase-space coordinates of each point on an
orbit can be reconstructed from the orbit's path across the sky and
measurements of the line-of-sight velocity along the orbit. In this paper we
complement this finding by showing that the orbit can also be reconstructed
if we know proper motions along the orbit rather than the radial velocities.
We also show that accurate proper motions of stream stars would enable
distances to be determined to points on the stream that are independent of
any assumption about the Galaxy's gravitational potential. Such ``Galactic
parallaxes'' would be as fundamental as conventional trigonometric
parallaxes, but measureable to distances $\sim70$ times further.

\end{abstract}


\section{Introduction}

The precision photometry for millions of faint stars observed by the Sloan
Digital Sky Survey (SDSS) led to the discovery of numerous tidal streams of
halo stars. Many of these streams are certainly of tidal origin because the
progenitor has been seen
\citep{Odenkirchen02,Majewski04,N5466,Fellhauer07,Grillmair} but in some
other cases the progenitor is unknown and may no longer be extant
\citep{Grillmair06,Fieldstars,GD-1}. It has long been recognised that streams
provide an important diagnostic of the still uncertain Galactic gravitational
field by virtue of the closeness with which a thin stream approximates an
orbit \citep{JohnstonHB}. However, the traditional way of exploiting this
connection, which is to search for orbits that are consistent with the data,
has yielded fewer convincing fits to the data than one might have expected,
and in any given case it is not clear why a better-fitting orbit has not been
found. 

Recently it was realised that given a Galactic potential $\Phi$ and
line-of-sight velocities along a stream, one can uniquely solve for the six
phase-space coordinates that points on the stream must have if they are to
trace an orbit in the given potential \citep[][hereafter
Paper I]{complexA,Binney08}. If the wrong gravitational potential is used in
the reconstruction, the recovered phase-space coordinates will in general be
inconsistent with conservation of energy (Paper I) and will violate the
tangential component of the equations of motion \citep[][hereafter Paper
II]{EyreB09}. Hence the reconstruction technique provides a powerful
diagnostic of the gravitational potential, and once the potential has been
determined, it will provide distances to stars that lie on streams that are
as absolute as trigonometric parallaxes but very much more precise than will
be possible for such distant objects in the foreseeable future (Paper I).

The main obstacles to attainment of these exciting goals are (a) the fact
that streams differ slightly but significantly from orbits, and (b) a lack of
reliable line-of-sight velocities along streams. Paper II addresses problem
(a). This paper addresses problem (b) by showing that proper motions may be
employed instead of line-of-sight velocities.

Many of the most promising streams have distances in the range $10-50\kpc$,
so their distance moduli are $15-18.5$ and their solar-type stars have
apparent magnitudes in the range $I\simeq19-22.5$. Perhaps the closest
streams of interest are the GD-1 and Anticentre streams
\citep{GD-1,Grillmair06}, which are only $\sim10\kpc$ distant.  Consequently
at $r<19$ \cite{Koposov} were able to obtain velocities for 24 turnoff stars
in the GD-1 stream, while at $g<20$ \cite{anticentre} measured velocities for
$\sim20$ stream stars in each of two fields. The situation regarding
velocities of stars in the more distant Orphan stream is much less
satisfactory -- \cite{orphans} conclude that indications of the line-of-sight
velocity of the Orphan stream ``are suggestive rather than conclusive''.
Even with an $8\,$m telescope it is extremely challenging to measure the
line-of-sight velocities to a few $\!\kms$ of significant numbers of
main-sequence stars at distances in excess of $20\kpc$. Consequently, the
strategy generally adopted with more distant streams is to identify giant
stars that probably belong to the stream and measure their velocities. For
example, \cite{Odenkirchen09} used the VLT to measure the line-of-sight
velocities of 74 giant stars with $i<18.4$ in the region of the Pal 5 stream
and concluded that only 17 of these stars were stream members; because the
stream is defined by main-sequence stars, not giants, one cannot be sure
that a giant observed in the direction of the stream is not a foreground or
background object.  Moreover, the number of main-sequence stars in a length
of stream is large compared to the number of giants, so there is much greater
scope for beating down random errors if main-sequence stars can be used. 

Since streams are identified from the photometry of individual main-sequence
stars, it is in principle possible to measure proper motions for all the stars
that define the stream. Such work is already possible  with the
SDSS survey \citep{Munnetal}, and work of significantly greater precision will be possible
with the Pan-Starrs survey, which is currently getting underway. In this
paper we show that orbit reconstruction is possible given proper motions
rather than line-of-sight velocities.

In Section 2 we describe the algorithm. Section 3 introduces the concept of
Galactic parallaxes which arises in connection with the algorithm.  Section 4
reports tests of the algorithm. Section 5 sums up and looks ahead.

\section{The algorithm}

We work in the inertial coordinate system in which the Galactic centre
is at rest. Let $\vect{r}_0$ be the position vector of
the Sun, $\vect{r}$ that of a star in the stream and let
$\vect{s}$ be the vector from the Sun to the star:
\begin{equation}
\vect{s} = s \vecthat{s} = \vect{r} - \vect{r}_0,
\label{eq:vectors}
\end{equation}
 where $\vecthat{s}$ is the direction from the Sun to the star. We must
distinguish between the derivative $\D\vecthat{s}/\D t$ that takes into
account the velocity $\vect{v}_0$ of the Sun, and the derivative
$\d\vecthat{s}/\d t$ that does not; the latter is tangent to the stream
and is  the proper motion that
would be measured by an observer who is stationary at the current location of
the Sun, while the former is the observable proper
motion of a star and has a component perpendicular to the stream.
If $u$ is the angle along the
stream from some fiducial point, then we may write
\begin{equation}
\frac{\D\vecthat{s}}{\D t} = \mu \vecthat{t}\quad ;\quad
\frac{\d\vecthat{s}}{\d t} = \dot{u}
\vecthat{p},\label{eq:derivatives}
\end{equation}
 where $\mu$ and $\vecthat{t}$ are the magnitude and direction of the
measured proper motion, while $\dot{u}$ and $\vecthat{p}$ are the
magnitude and direction of the motion along the stream. 

The space velocities measured by an observer moving with  the Sun and one
stationary at the Sun's location are related by
 \beq
{\D\vect{s}\over\D t}={\d \vect{s}\over\d t}-\vect{v}_0.
\eeq
 With equations  \eqref{eq:derivatives} we can therefore write
\begin{equation}
{\D s\over \D t}\vecthat{s} + s\mu \vecthat{t} = {\d s\over\d t} \vecthat{s} +
s \dot{u} \vecthat{p} - \vect{v}_0,
\end{equation}
 where $\D s/\D t$ is the spectroscopically measured heliocentric velocity
and $\d s/\d t=v_\parallel$ is the projection along the line of sight of the
star's velocity with respect to the Galactic centre.  Equating components in
the plane of the sky, we have
 \begin{equation}
s \mu \vecthat{t} = s \dot{u} \vecthat{p} - (\vect{v}_0 - \vecthat{s} \cdot 
\vect{v}_0\, \vecthat{s} ) = s \dot{u} \vecthat{p} - \vect{v}_{0\perp},
\end{equation}
 where $\vect{v}_{0\perp}$ is the component of the Sun's velocity perpendicular to the
line of sight. This equation has just two unknowns, $\dot{u}$ and $s$, and we
can in principle solve for both through
\begin{equation}
\dot{u}\vecthat{p} = \mu \vecthat{t} + \frac{\vect{v}_{0\perp}}{s}.
\label{eq:fundamental}
\end{equation}
 Specifically, since both $\vecthat{t}$ and $\vecthat{p}$ can in principle be
deduced from the observations and $\vect{v}_{0\perp}$ may be presumed known,
we could determine $s$ such that the right side is parallel to $\vecthat{p}$, and
then read off $\dot{u}$ from the magnitude of the right side. Unfortunately,
the uncertainty in the direction $\vecthat{t}$ is likely to be
significant. We therefore eliminate it by squaring up,
\begin{equation}
(s\dot{u})^2 - 2{\vect{v}_{0\perp} \cdot \vecthat{p}}\,s\dot{u}
+ {\left| \vect{v}_{0\perp} \right|^2} - (s\mu)^2 = 0.
\end{equation}
The roots of this quadratic equation for $s\dot u$ are
\begin{equation}
s \dot{u} = \vect{v}_{0\perp} \cdot \vecthat{p} \pm \sqrt{ 
   (\vect{v}_{0\perp} \cdot \vecthat{p})^2  - 
   \left| \vect{v}_{0\perp} \right|^2+ (s \mu)^2}.\label{eq:udot}
\end{equation}
 Using this expression to eliminate $\dot u$ from equation
\eqref{eq:fundamental}, we find
 \beq
\pm\sqrt{ 
   (\vect{v}_{0\perp} \cdot \vecthat{p})^2  - 
   \left| \vect{v}_{0\perp} \right|^2+ (s
   \mu)^2}=\mu\,\vecthat{t}\cdot\vecthat{p}.
\eeq
 Since $\mu$ is inherently positive, the sign in front of the radical in
equation \eqref{eq:udot} must be chosen to agree with the sign of
$\vecthat{t}\cdot\vecthat{p}$. Even though the directions of individual
proper motions may be uncertain, it should be possible to decide whether they
are on average opposed to the direction of travel along the stream.
Equation (\ref{eq:udot}) makes $\dot u$ into a
function of $s$ and quantities that can be determined from the observations.

Now let $\b F(\b r)$ be the Galaxy's gravitational acceleration ($\b
F=-\nabla\Phi$). We recall that when we resolve the star's equation of
motion along the line of sight to the Sun, we obtain (Paper~I)
\beq
\label{eq:eq-of-m}
{\d\vlos\over\d t}= \dot{u} {\d\vlos \over \d u} = \Flos+{\vperp^2\over s},
\eeq
where the subscripts $\parallel$ and $\perp$ denote components along
and perpendicular to $\vecthat{s}$, respectively.

Since $\dot u=v_\perp/s$ and equation \eqref{eq:udot} makes $\dot u$ a known
function of  $s$, with equation (\ref{eq:eq-of-m}) we can now write down
a system of three differential equations for the
unknowns along the stream:
 \begin{eqnarray}\label{eq:diff-eqs}
{\d s\over\d u}&=&\vlos \over \dot{u}\nonumber\\
{\d\vlos\over\d u}&=&{\Flos+s\dot u^2 \over \dot{u}}\\
{\d t \over \d u} & = & {1 \over \dot{u}}\,.\nonumber
\end{eqnarray}
 To integrate these equations along the stream we need initial conditions for
$s$, $v_\parallel$ and $t$. We can trivially set $t=0$ at a fiducial point of
the stream and guess a value $s_0$ for $s$ at that point. Then we can compute
the initial value of $v_\parallel$ as follows.

We write
\begin{eqnarray}\label{eq:Fperp}
F_\perp&=&{\d\b v\over\d t}\cdot\hat\b p
={\d\over\d t}(\vlos\vecthat{s}+s\dot u\hat\b p)\cdot\hat\b p\nonumber\\
&=&\vlos{\d\vecthat{s}\over\d t}\cdot\hat\b p+\vlos\dot u+s\ddot u\\
&=&2v_\parallel\dot u+s\ddot u,\nonumber
\end{eqnarray}
 where we have used equation \eqref{eq:derivatives} to eliminate
 $\d\vecthat{s}/\d t$. Thus
 \beq
\frac{\d(s \dot{u})}{\d t} = \dot{u}{\d (s \dot{u}) \over \d u}
= F_\perp - \vlos \dot{u}. \label{eq:tan-eq-of-m}
\eeq
 The left side of this equation is obtained by explicitly differentiating
equation (\ref{eq:udot}),
 \begin{equation}
\frac{\d (s\dot{u})}{\d u}  = \alpha
+ {1 \over \beta} \left( \alpha\gamma
 - \vot \cdot \frac{\d\vot}{\d u}+ \mu s^2 \frac{\d\mu}{\d u}
+ {s \mu^2 v_\parallel \over \dot{u}}
\right),  \label{eq:dsudot/dt}
\end{equation}
 where we have defined,
\begin{eqnarray}
\alpha &=& \frac{\d\vot}{\d u}\cdot\vecthat{p}
+ \vot \cdot \frac{\d\vecthat{p}}{\d u},\nonumber\\
\beta &=& \sqrt{(\vot \cdot \vecthat{p})^2
- \left|\vect{v}_{0\perp} \right|^2 + (s\mu)^2 },\\
\gamma &=& \vot \cdot \vecthat{p}.\nonumber
\end{eqnarray}
 Equation (\ref{eq:dsudot/dt}) is linear in $v_\parallel$ and with equation
 (\ref{eq:tan-eq-of-m}) it readily
 yields
\begin{eqnarray}
v_\parallel = \Bigl({\beta F_\perp \over \dot{u}} - \alpha\beta - \alpha\gamma
  + \vot &\cdot&\frac{\d\vot}{\d u}- \mu {\d\mu\over\d u}s^2\Bigr)\nonumber\\
&&\Big/\left({\beta + s \mu^2/\dot u }\right).
\end{eqnarray}
 Once the distance $s_0$ to the fiducial point has been chosen, the right
 side of this equation can be evaluated from the data because differentiating
$\vect{v}_{0\perp}=\vect{v}_0-\vecthat{s}\cdot\vect{v}_0\,\vecthat{s}$ along
the stream yields
 \begin{eqnarray}
\frac{\d\vot}{\d u} & = &
- \frac{\d\vecthat{s}}{\d u}\cdot\vect{v}_0\,\vecthat{s}
 -\vecthat{s}\cdot\vect{v}_0\,\frac{\d\vecthat{s}}{\d u} \nonumber\\
& = &- \vecthat{p}\cdot\vect{v}_0\,\vecthat{s}  
-\vecthat{s}\cdot\vect{v}_0\,\vecthat{p}.
\end{eqnarray}
 Hence
\begin{equation}
\frac{\d\vot}{\d u}\cdot\vecthat{p} = -\vecthat{s}\cdot\vect{v}_0,
\end{equation}
and
\begin{equation}
\vot\cdot\frac{\d\vot}{\d u} = -(\vecthat{s}\cdot\vect{v}_0)\, \vot \cdot
\vecthat{p}.
\end{equation}
 Thus the initial conditions required for the integration of equations
(\ref{eq:diff-eqs}) follow once $s_0$ has been chosen. The solution to these
equations completes the information required to assign a full six-dimensional
phase-space position to every point on the stream.

\begin{figure}
\centerline{\epsfig{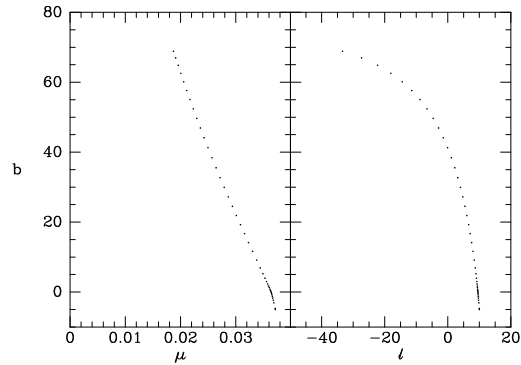}}
\caption{Right panel: the projection onto the sky of a numerically integrated
orbit in a Myamoto--Nagai potential with $b/a=0.2$. Left panel: the proper
motion as a function of $b$.}\label{fig:lbplot}
\end{figure}

\begin{figure}
\centerline{\epsfig{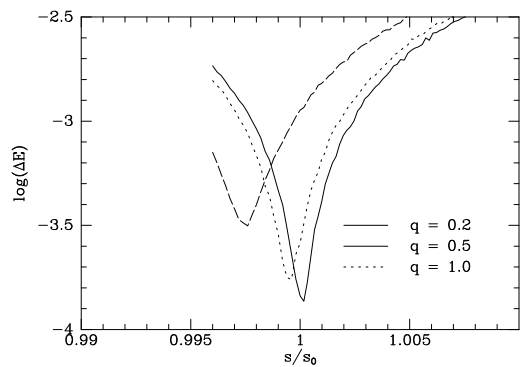}}
\caption{The log to base 10 of the rms variation in the energy when the orbit
shown in \figref{fig:lbplot} is reconstructed from an assumed fiducial distance
$s$ rather than its true value $s_0$. The triangles show results obtained
when the reconstruction employs the true potential, while the squares and
pentagons are for less-flattened potential.}\label{fig:deplot}
\end{figure}

\section{Galactic parallax}

If we knew $\vect{v}_0$ accurately, we could determine the distance to a
stationary star from the magnitude of its proper motion, which would be
entirely due to the Sun's motion. Similarly, in so far as we can argue that a
star in a stream has no velocity perpendicular to the stream, we can
determine the distance to the star by attributing to the Sun's motion the
star's proper motion perpendicular to the stream -- equation
\eqref{eq:fundamental} embodies this idea mathematically. Distances obtained
in this way without the use of dynamics would have the same logical status as
conventional trigonometric parallaxes, and might be called ``Galactic
parallaxes''. For given astrometric precision Galactic parallaxes could be
accurately measured to much greater distances than conventional parallaxes
because in three years the Sun moves $\sim140\,$AU around the Galaxy, leading
to a change in the position of an object that is $\sim70$ times larger than
the corresponding conventional parallax angle. Consequently equipment such as
Gaia that can measure the conventional parallaxes of sources at distances of
order $10\kpc$ could measure Galactic parallaxes out to $700\kpc$, i.e., as
far as the Andromeda galaxy. Unfortunately, before this method could be
applied to streams within M31, one would have to determine the velocity of
the Sun relative to the barycentre of M31, which we are not likely to know
better than we know the distance to the centre of M31.

By measuring the Galactic parallax to each point along a stream, the stream's
three-dimensional trajectory could be determined without any knowledge of the
Galaxy's gravitational potential.  The requirement that this trajectory be an
orbit must constrain the potential rather tightly.

In the previous section we have chosen to sacrifice some of the diagnostic
power of equation \eqref{eq:fundamental} by eliminating $\vecthat{t}$ on the
grounds that it will be hard to measure. With $\vecthat{t}$ eliminated, the
distance can only be recovered by adopting a trial gravitational potential
and searching over the fiducial distance $s_0$.

For sufficiently small $s$, the radical in equation \eqref{eq:udot} becomes
imaginary, so there is a lower bound on the values of $s_0$ that should be
considered: below this bound the values of $s$ obtained by solving equations
\eqref{eq:diff-eqs} will somewhere approach the value at which the radical in
equation  \eqref{eq:udot} becomes imaginary. This event signals that the measured
value of $\mu$ is too small to be consistent with the reflex of the Sun's
velocity at the proposed distance. Thus the kinematics of the problem imposes
a lower bound on $s_0$. There is no similar kinematic upper bound on $s_0$
because the radical in equation \eqref{eq:udot} is real for all large $s$.

\section{Tests}

We have tested the ability of the  algorithm to reconstruct orbits in the
same Miyamoto-Nagai (1975) potential that was used in Paper I, namely
\beq
\Phi(R,z)=-{GM\over\sqrt{R^2+\bigl(a+\sqrt{z^2+q^2a^2}\bigr)^2}}
\eeq
 with $q=0.2$. The numerical procedures were essentially the same as those
described in Paper I except for one significant item: the fiducial point was
placed in the middle of the stream rather than at one of its ends, and 
equations
\eqref{eq:diff-eqs} were integrated in both directions away from this point.
This modification is advantageous as the data constrain derivatives of the
observables much more tightly at the middle of the stream than at its ends.

The right panel of \figref{fig:lbplot} shows the track over the sky of an
orbit that starts at galactic latitude $b=-5^\circ$ and a distance $s_0=15a$
as viewed from the Sun, which is on a circular orbit at $R_0=8a$.  The left
panel shows the magnitude of the proper motion along the orbit. The points in
\figref{fig:deplot} show the rms variation in the energy of orbits that are
reconstructed from the 42 data points shown in \figref{fig:lbplot} for
various assumed distances $s$ to the fiducial point and three trial
potentials: the true potential, which has scale-length ratio $q=0.2$ and two
less flattened potentials. The rms variation in $E$ has a
sharp minimum at the true distance when the true potential is used, and
higher minima when the wrong potential is used. The exquisite precision with
which the distance can be determined from this plot is remarkable: better
than three parts in $10^4$. 

In \figref{fig:deplot} the smallest value reached by the rms energy variation
is significantly higher than the corresponding figure for the same orbit when
line-of-sight velocities are used (Paper I). However, for other orbits
smaller values of the rms variation in energy are obtained from  proper-motion
data; not surprisingly the data with the greatest  diagnostic power varies
with the nature of the orbit.

\section{Conclusions}

We have complemented the work of Paper I by showing that when proper motions
can be measured along a section of a single orbit, the full phase-space
coordinates for the orbit can be reconstructed as readily as is the case when
line-of-sight velocities have been measured.  However, we have also recovered
a more powerful result: if the direction as well as the magnitude of the
proper motions can be accurately measured, the three-dimensional geometry of
the orbit can be recovered without assuming anything about the Galaxy's
gravitational potential. This reconstruction is possible because the
Sun's velocity must be responsible for the motion of stars perpendicular to
the orbit, so from proper motion perpendicular to an orbit distances can be
inferred that are as fundamental as conventional trigonometric parallaxes. 

The extension of the work of Paper I to proper motions is useful because
proper motions can be readily measured for the main-sequence stars that
define the stream.  These stars are typically so faint ($I\gta20$) that it is
hard to obtain line-of-sight velocities of the requisite accuracy for
large numbers of them.

In Paper II we show that even when line-of-sight velocities are available, a
combination of observational errors and the fact that a tidal stream does not
strictly follow an orbit make the task of determining whether a given
potential is compatible with an observed stream complex. It proves necessary
to search a multi-dimensional space of phase-space tracks that are compatible
with the data for ones that are dynamically consistent orbits. A comparable
search would be required when the data included proper motions rather than
line-of-sight velocities. In a future paper we will extend to proper
motions the techniques for handling this problem that are developed for
line-of-sight velocities in Paper II.

Another interesting avenue is to determine the solar motion by
treating it as an unknown when reconstructing orbits for several streams,
using either line-of-sight velocities or proper motions, and choosing
the value which allows a consistent interpretation of these
streams. Recently \cite{Koposov} used a combination of line-of-sight
velocities and proper motions for the GD-1 stream to constrain $\vect{v}_0$.
Stronger constraints should be attainable by modelling several streams
simultaneously.

\section*{Acknowledgments}

We thank the anonymous referee for his/her suggestions. AE acknowledges
the support of PPARC/STFC during preparation of this work.

\label{lastpage}

\end{document}